\documentclass[final,3p,times]{elsarticle}

\usepackage{graphicx}

\usepackage{amssymb}
\usepackage{amsthm}

\usepackage{hyperref}
\usepackage{verbatim}

 \biboptions{square,sort&compress,comma}

\journal{Computer Physics Communications}

\usepackage{mathrsfs}
\usepackage{amssymb}
\usepackage{amsmath}
\usepackage{bm}
\usepackage{graphicx}
\usepackage{color}

\begin{document}

\begin{frontmatter}



\title{PDRK: A General Kinetic Dispersion Relation Solver for Magnetized Plasma}

\date{\today}


\author{Hua-sheng XIE}
\ead{huashengxie@gmail.com}

\author{Yong Xiao}
\address{Institute for
Fusion Theory and Simulation and the Department of Physics, Zhejiang
University, Hangzhou, 310027, People's Republic of China}

\begin{abstract}
A general, fast, and effective approach is developed for numerical
calculation of kinetic plasma dispersion relations. The plasma
dispersion function is approximated by $J$-pole expansion.
Subsequently, the dispersion relation is transformed to a standard
matrix eigenvalue problem of an equivalent linear system. The result
is accurate for $J=8$ except the solutions that are the little
interesting heavily damped modes. In contrast to conventional
approaches, such as Newton's iterative method, this approach can
give either all the solutions in the system or a few solutions
around the initial guess. It is also free from convergent problems.
The approach is demonstrated from electrostatic one-dimensional and
three-dimensional dispersion relations, to electromagnetic kinetic
magnetized plasma dispersion relation
for bi-Maxwellian distribution with parallel velocity drift.\\
\bigskip \\
\textbf{Program summary}\\
\\
\emph{Title of program:} PDRK \\
\emph{Catalogue identifier:}   \\
\emph{Program summary URL:}  \\
\emph{Program obtainable from:} CPC Program Library, Queen
University of Belfast, N.
 Ireland\\
\emph{Computer for which the program is designed and others on
which it
has been tested: Computers:} Any computer running \emph{MATLAB} 7. Tested on Lenovo T430.\\
\emph{Operating systems under which the program has been tested:} Windows 8\\
\emph{Programming language used:} \emph{MATLAB} 7
\\
\emph{Memory required to execute with typical data:} 500 M\\
\emph{No. of lines in distributed program, including test data,
etc.:} 700 \\
\emph{No. of bytes in distributed program, including test data,
etc.:} 30 000 \\
\emph{Distribution format:} .tar.gz \\
\emph{Nature of physical problem:} Solving kinetic dispersion relations for multi-species plasmas. \\
\emph{Method of solution:} Transforming to an equivalent linear system and then solving as matrix eigenvalue problem.\\
\emph{Restrictions on the complexity of the problem:} Not suitable for heavily damped modes and only non-relativistic version at present.\\
\emph{Typical running time:} About 1 minutes on a Intel 2.60 GHz PC. \\
\emph{Unusual features of the program:} Can give all interesting
solutions fastly and without convergent difficulty.

\end{abstract}


\end{frontmatter}


\section{Introduction}\label{sec:intro}
Given the richness of waves and instabilities in astrophysical,
space, laser, and laboratory plasmas, studying the corresponding
linear dispersion relations of different plasma systems is of
practical interest. However, except for some simple cases, the
dispersion relations are usually too complicated to be solved either
analytically or even numerically.

The multi-fluid plasma dispersion relation has been numerically
solved generally using matrix method in a previous work, i.e.,
PDRF\cite{Xie2014}.

At present, several multi-component magnetized kinetic plasma
dispersion relations solvers are available, such as WHAMP by
Ronnmark\cite{Ronnmark1982,Ronnmark1983}, NHDS by Verscharen {\it et
al.} \cite{Verscharen2013}, and solvers by Gary {\it et al.}
\cite{Gary1993,Gary2011}, by Willes and
Cairns\cite{Willes2000,Verdon2009} and by Lin {\it et
al.}\cite{Lin2005}, among others. However, all these solvers obtain
the dispersion relations from the determinant of the corresponding
3-by-3 dielectric tensor using a given initial guess. These solvers
are usually time consuming and have difficulty showing a complete
picture of the modes in the system. Furthermore, these solvers may
also suffer from convergence problems because the plasma dispersion
function $Z(\zeta)$ and Bessel functions (especially in high-order
cyclotron frequencies, e.g., $\omega>10\Omega_c$, where $\Omega_c$
is the cyclotron frequency) have several solutions around a given
frequency. Thus, a careful selection of the initial guess is
required to make it converge to the solution we want.

In this work, we extend our previous work, a multi-fluid dispersion
relation solver\cite{Xie2014}, to a general kinetic version, but
still maintain the use of a full-matrix approach. In contrast, two
additional steps are required in the kinetic version: solving for
the plasma dispersion function $Z(\zeta)$ and seeking an equivalent
linear system. The first step is accomplished by $J$-pole expansion
(Pad\'e approximation) as used by Martin {\it et
al.}\cite{Martin1980} and Ronnmark\cite{Ronnmark1982,Ronnmark1983}.
The first step has also been used by Cereceda and
Puerta\cite{Cereceda2000} to solve the electrostatic 1D (ES1D)
system. Physical interpretations of the Pad\'e approximation of
$Z(\zeta)$ are given by Tjulin {\it et al.}\cite{Tjulin2000} and
Robinson and Newman\cite{Robinson1988}. The second step is more
difficult and should be treated on a case-to-case basis as we can
see in the following sections.

\section{Electrostatic systems}\label{sec:es}
We start with simple electrostatic systems to show how our approach
can be implemented.

\subsection{Electrostatic 1D}\label{sec:es1d}
First, we solve the simplest multi-component electrostatic 1D (ES1D)
problem with drift Maxwellian distribution $f_{s0} =
(\frac{m_s}{2\pi k_B T_s})^{1/2}
\exp[-\frac{(v-v_{s0})^2}{2k_BT_s}]$. The dispersion relation is
\begin{equation}\label{eq:drkes1d}
    D=1+\sum_{s=1}^S\frac{1}{(k\lambda_{Ds})^2}[1+\zeta_sZ(\zeta_s)]=0,
\end{equation}
where $\lambda_{Ds}^2=\frac{\epsilon_0k_BT_s}{n_sq_s^2}$,
$v_{ts}=\sqrt{\frac{2k_BT_s}{m_s}}$ and
$\zeta_s=\frac{\omega-kv_{s0}}{kv_{ts}}$. Unmentioned notations are
standard. The plasma dispersion function can be approximated using
$J$-pole expansion
\begin{equation}\label{eq:NZ}
    Z(\zeta)\simeq Z_J(\zeta)=\sum_{j=1}^J\frac{b_j}{\zeta-c_j},
\end{equation}
where $J=8$ is used by Ronnmark \cite{Ronnmark1982,Ronnmark1983} and
$J=2,3,4$ are provided by Martin {\it et al.}\cite{Martin1980},
producing accurate results for most domains (except
$y<\sqrt{\pi}x^2e^{-x^2}$ when $x\gg1$, with $\zeta=x+iy$),
especially in the upper plane. However, the method does not perform
well for heavily damped modes, which are of little interest anyway.
For completeness, the coefficients $c_j$ and $b_j$ for $J=4$, $J=8$
and $J=12$ (see \ref{sec:jexpan}) are provided in Table
\ref{tab:NpoleZ}. Note the useful relations $\sum_jb_j=-1$,
$\sum_jb_jc_j=0$ and $\sum_jb_jc_j^2=-1/2$.

\begin{table}
\begin{center}
\caption{\label{tab:NpoleZ} The coefficients  $c_j$ and $b_j$ for
$J=4$\cite{Martin1980}, $J=8$\cite{Ronnmark1982} and $J=12$
(\ref{sec:jexpan}) under $J$-pole approximations of $Z(\zeta)$,
where the asterisk denotes complex conjugation.}
\begin{tabular}{ccc} \hline\hline
  & $b_1$=0.546796859834032 + 0.037196505239277i & $c_1$=1.23588765343592 - 1.21498213255731i \\
   $J=4$ & $b_2$=-1.046796859834027 + 2.101852568038518i & $c_2$=-0.378611612386277 - 1.350943585432730i\\
  & $b(3:4)$=$b^*(1:2)$ &  $c(3:4)$=$-c^*(1:2)$ \\\hline
  & $b_1$=-1.734012457471826E-2-4.630639291680322E-2i & $c_1$=2.237687789201900-1.625940856173727i \\
  & $b_2$=-7.399169923225014E-1+8.395179978099844E-1i & $c_2$=1.465234126106004-1.789620129162444i \\
   $J=8$ & $b_3$=5.840628642184073+9.536009057643667E-1i & $c_3$=0.8392539817232638-1.891995045765206i\\
  & $b_4$=-5.583371525286853-1.120854319126599E1i & $c_4$=0.2739362226285564-1.941786875844713i \\
  & $b(5:8)$=$b^*(1:4)$ &  $c(5:8)$=$-c^*(1:4)$ \\\hline
  & $b_1$=-0.004547861216840 + 0.000621096229879i & $c_1$=2.978429162453205 - 2.049696666440972i \\
  & $b_2$=0.215155729059403 - 0.201505401705763i & $c_2$=-2.256783783969929 - 2.208618411911446i \\
  & $b_3$=0.439545043457674 - 4.161084685092405i & $c_3$=1.673799856114519 - 2.324085194217706i\\
   $J=12$ & $b_4$=-20.216967308177410 + 12.885503528244977i & $c_4$=1.159032034062764 - 2.406739409567887i \\
  & $b_5$=67.081488119986460 - 20.846345891864550i & $c_5$=-0.682287637027822 - 2.460365014999888i \\
  & $b_6$=-4.801467372237129e+01 - 1.072756140299431e+02i & $c_6$=0.225365375295874 - 2.486779417872603i \\
  & $b(7:12)$=$b^*(1:6)$ &  $c(7:12)$=$-c^*(1:6)$ \\\hline\hline
\end{tabular}
\end{center}
\end{table}

Combining (\ref{eq:drkes1d}) and (\ref{eq:NZ}), yields
\begin{equation}\label{eq:drkes1d2}
    1+\sum_s\sum_j\frac{b_{sj}}{(\omega-c_{sj})}=0,
\end{equation}
with $b_{sj}=\frac{b_jc_jv_{ts}}{k\lambda_{Ds}^2}$ and
$c_{sj}=k(v_{s0}+v_{ts}c_j)$. An equivalent linear system can be
obtained as follows:
\begin{subequations} \label{eq:fpeq}
\begin{eqnarray}
  & \omega n_{sj} = c_{sj}n_{sj}+b_{sj}E,\\
  & E = -\sum_{sj}n_{sj},
\end{eqnarray}
\end{subequations}
which is an eigenvalue problem of a $SJ\times SJ$ dimensional eigen
matrix $\bm M$, i.e., $\omega X=\bm M\bm X$, with $SJ=S\times J$ and
$\bm X=\{n_{sj}\}$. The singularity in the denominator of
(\ref{eq:drkes1d2}), which is encountered in conventional methods,
can be canceled by using the transformation (\ref{eq:fpeq}). Hence,
the matrix method can easily support multi-component systems.

\begin{figure}
\begin{center}
  \includegraphics[width=10cm]{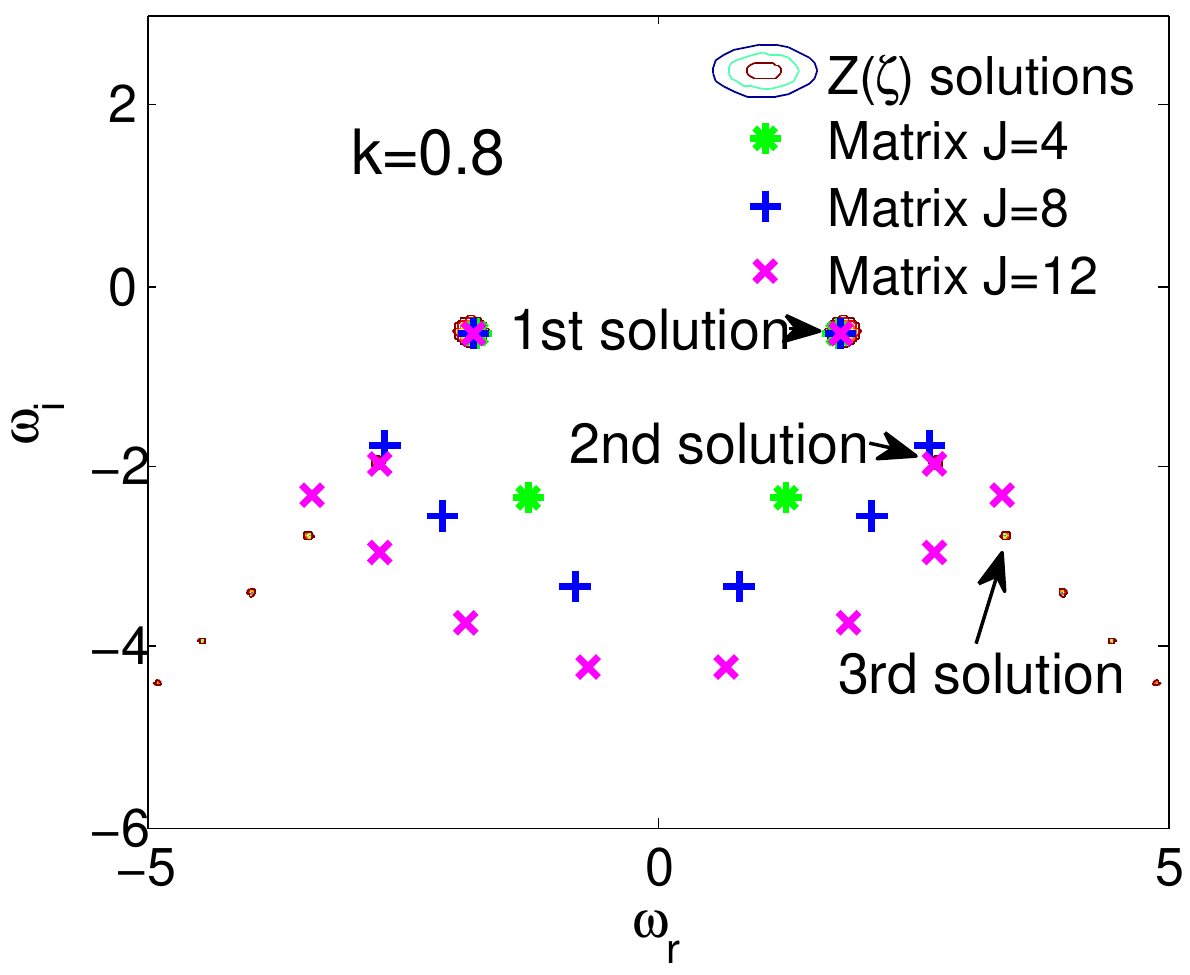}\\
  \caption{Comparison of all the solutions obtained from the matrix method and the $Z(\zeta)$ function.}\label{fig:ZvsJpole}
\end{center}
\end{figure}

For Langmuir wave Landau damping, calculating the largest imaginary
part solution using matrix method ($\omega^M$) and the original
$Z(\zeta)$ function ($\omega^Z$)\cite{Xie2013} are shown in Table
\ref{tab:LD}. We can see that the result of the matrix method is
accurate in $10^{-4}$ when $J=8$ and the error for $J=4$ is also
small (10\%). Thus, we have verified that our approach is feasible.
In principle, infinite numbers of frequency solutions exist for a
fixed wave vector $k$ (the physical discussions can be found in
Ref.\cite{Xie2013a} and references in). Fig.\ref{fig:ZvsJpole} shows
all the solutions of the matrix method and the solutions using
$Z(\zeta)$ function for $k\lambda_{De}=0.8$. The largest imaginary
part solutions (first solution) are almost identical, which is our
objective. However, other heavily damped solutions should be
excluded due to the poor approximation in those ranges. For example,
the error for the second solution between the $Z(\zeta)$ solution
and the $J=8$ solution is around 10\%, whereas the third solution is
completely wrong for $J=8$. Fortunately, for most studies, these
heavily damped modes are of little interest. The $J=12$ results can
be more accurate ($10^{-7}$) as shown in Table \ref{tab:LD} and
Fig.\ref{fig:ZvsJpole}. In principle, Eq.(\ref{eq:drkes1d}) has no
singularity for $k\neq0$. Given the existence of multiple solutions,
if the initial guess is not good, then root finding cannot converge
to the desired solutions.

\begin{table}
\begin{center}
\caption{\label{tab:LD} Comparison of the Landau damping solutions
using the matrix method and the original $Z(\zeta)$ function. Here,
$\omega$ is normalized by
$\omega_{pe}=\sqrt{{n_ee^2}/{\epsilon_0m_e}}$.}
\begin{tabular}{ccccccccc} \hline\hline
  $k\lambda_{De}$ & $\omega_r^M(J=4)$ & $\omega_i^M(J=4)$ & $\omega_r^M(J=8)$ & $\omega_i^M(J=8)$ & $\omega_r^M(J=12)$ & $\omega_i^M(J=12)$ & $\omega_r^Z$ & $\omega_i^Z$ \\\hline
  0.1 & 0.9956 &  9.5E-3 & 1.0152 & 1.7E-5 & 1.0152 & 9.5E-8 & 1.0152 & -4.8E-15 \\
  0.5 & 1.4235 & -0.1699 & 1.4156 & - 0.1534 & 1.4157 & -0.1534 & 1.4157 & -0.1534 \\
  1.0 & 2.0170 & -0.8439 &  2.0459 & - 0.8514 & 2.0458 & -0.8513 & 2.0458 & -0.8513 \\
  2.0 & 3.2948 & - 2.6741 & 3.1893 & - 2.8272 & 3.1891 & -2.8272 & 3.1891 & -2.8272 \\ \hline\hline
\end{tabular}
\end{center}
\end{table}

\begin{figure}
\begin{center}
  \includegraphics[width=15cm]{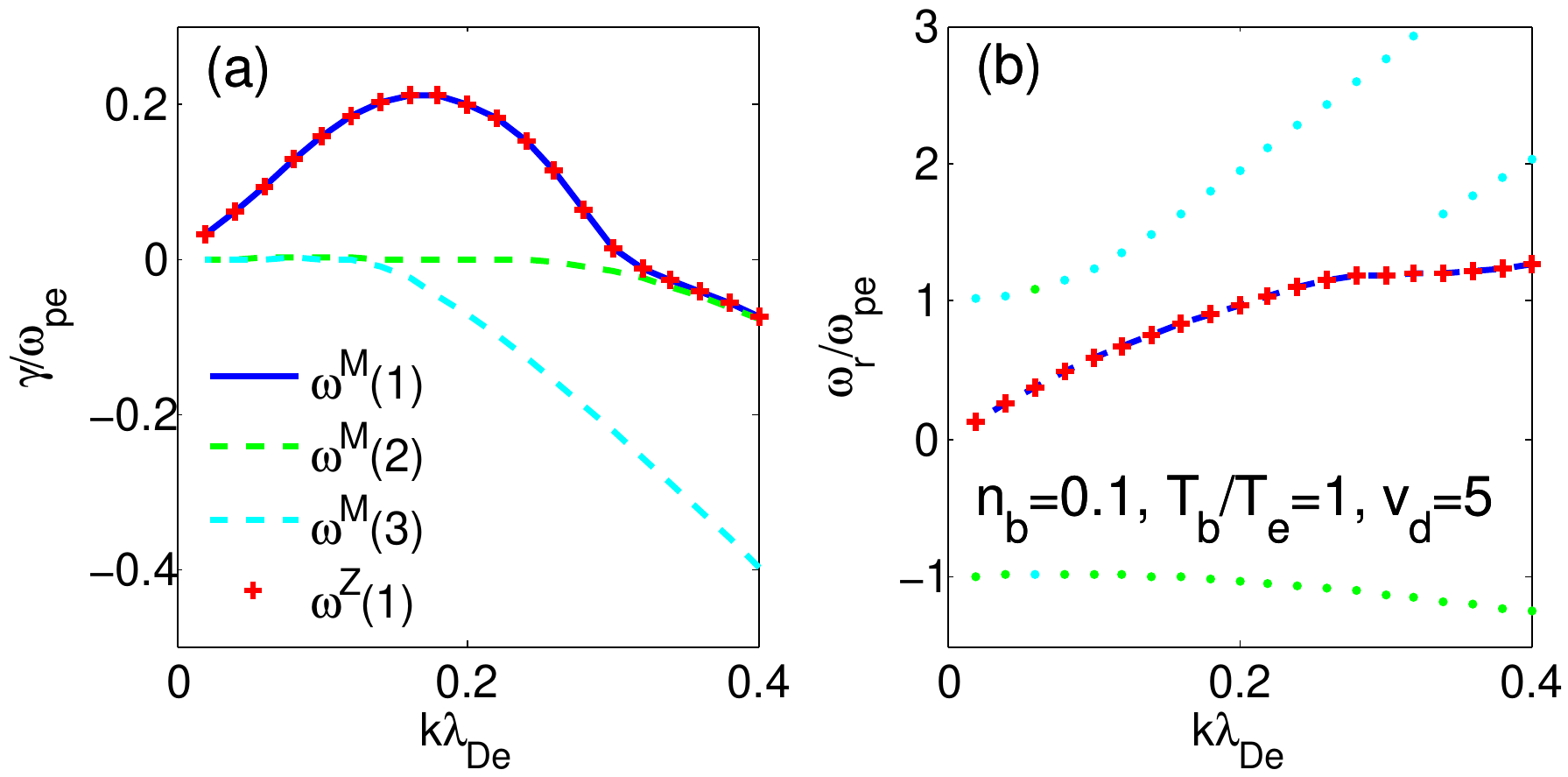}\\
  \caption{Comparison of the first three ($\omega^M$) largest imaginary part solutions obtained from the matrix method ($J=8$) and one
  solution ($\omega^Z$) obtained from $Z(\zeta)$ function for the bump-on-tail parameters.}\label{fig:beam_es1d_k}
\end{center}
\end{figure}

For the two-frequency-scale ion acoustic mode, besides the Langmuir
mode $\omega=2.0459-0.8513i$, the largest imaginary part solution
obtained from the matrix method ($J=8$) is also consistent with the
solution obtained from the $Z(\zeta)$ function, e.g., $T_i=T_e$,
$m_i=1836m_e$, $k\lambda_{De}=1$, gives $\omega=0.0420-0.0269i$.
Hereafter, $J=8$ will be used as default.

We further check the electron bump-on-tail mode ($s=e,b$), with
$T_b=T_e$, $v_b=5v_{te}$ and $n_b=0.1n_0$ ($n_e=n_0-n_b$). Both
$J=8$ matrix method and root finding using $Z(\zeta)$ function give
the same largest imaginary part solution $\omega=0.9785 + 0.2000i$
for $k\lambda_{De}=0.2$. The $J=4$ matrix method gives
$\omega=0.9772 + 0.2076i$. Fig.\ref{fig:beam_es1d_k} shows $\omega$
and $\gamma$ vs. $k$ for the above parameters, where the first three
largest imaginary part solutions from the matrix method ($J=8$) and
one solution from $Z(\zeta)$ function are shown. $\omega^Z$ is
identical to $\omega^M$. However, different initial guesses should
be tested to find other solutions when we using the $Z(\zeta)$
function. By contrast, no initial guess is required when using the
matrix method. Therefore, with matrix method, no important solutions
are missed.

\subsection{Harris dispersion relation}\label{sec:es3d}
We go further to solve a more complicated example, including the
$n$-th ($n=-\infty$ to $\infty$) order cyclotron frequency, i.e.,
the electrostatic 3D-magnetized (ES3D) Harris dispersion
relation\cite{Gurnett2005}
\begin{equation}\label{eq:drkes3d}
    D=1+\sum_{s=1}^S\frac{1}{(k\lambda_{Ds})^2}[1+\frac{\omega-k_zv_{s0}-n\Omega_s+\lambda_Tn\Omega_s}{k_zv_{zts}}\sum_{n=-\infty}^{\infty}\Gamma_n(b_s)Z(\zeta_{sn})]=0,
\end{equation}
where, $\lambda_{Ds}^2=\frac{\epsilon_0k_BT_{zs}}{n_{s0}q_s^2}$,
$v_{ts}=\sqrt{\frac{2k_BT_s}{m_s}}$, $\lambda_T=T_z/T_\perp$,
$\zeta_{sn}=\frac{\omega-k_zv_{s0}-n\Omega_s}{k_zv_{zts}}$,
$\Gamma_n(b)=I_n(b)e^{-b}$, $b_s=k^2_\perp\rho^2_{cs}$,
$\rho_{cs}=\sqrt{\frac{ v^2_{\perp ts}}{\Omega_s}}$, $I_n$ is the
modified Bessel function, and the equilibrium distribution is
assumed to be drift bi-Maxwellian $f_{s0}
=f_\perp(v_\perp)f_z(v_z)$, with $f_\perp=\frac{m_s}{2\pi k_B
T_{s\perp}}\exp[-\frac{m_sv_\perp^2}{2k_BT_{s\perp}}]$ and
$f_z=(\frac{m_s}{2\pi k_B T_{s\perp}})^{1/2}
\exp[-\frac{m_s(v_\parallel-v_{s0})^2}{2k_BT_{sz}}]$. The background
magnetic field is assumed to be ${\bm B_0}=(0,0,B_0)$, and the wave
vector ${\bm k}=(k_x,0,k_z)=(k\sin\theta,0,k\cos\theta)$, which
gives $k_\perp=k_x$ and $k_\parallel=k_z$.

This dispersion relation contains infinite-order summation of Bessel
functions. However, Eq.(\ref{eq:drkes3d}) is very similar to
Eq.(\ref{eq:drkes1d}). Thus, the transformation to an equivalent
linear system/matrix is the same and straightforward. In the
computation, we only keep the first $N$ Bessel functions, i.e.,
$n=-N$ to $N$. The dimensions of the eigen matrix is  $SNJ\times
SNJ$, with $SNJ=S\times(2N+1)\times J$. The singularity for
$k_z\to0$ around $\omega-n\Omega_{cs}\to0$ in (\ref{eq:drkes3d}) is
removed after the transformation.

\subsubsection{Electron Bernstein modes}
\begin{figure}
\begin{center}
  \includegraphics[width=15cm]{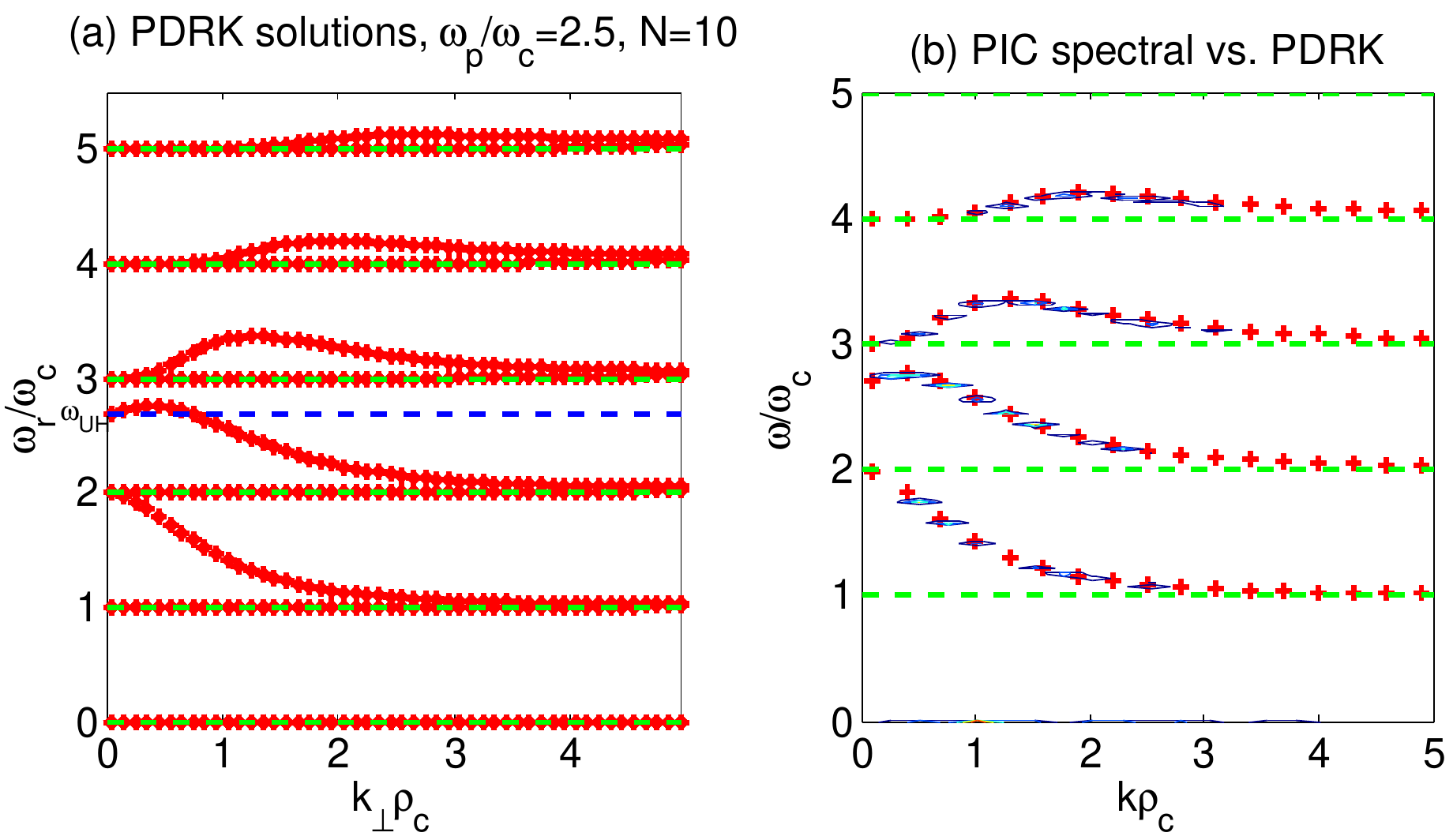}\\
  \caption{The electron Bernstein modes calculated from the Harris dispersion relation using the matrix
  method. The upper hybrid frequency calculated at the cold limit is $\omega_{UH}=\sqrt{\omega_c^2+\omega_p^2}=2.69$,
  which agrees with the matrix solution in the limit $k_{\perp}\rho_c\to0$. The plasma dispersion relation kinetic
  version (PDRK-ES3D) solutions also agree with the contour plot of the PIC spectra (b).}\label{fig:es3d_bern}
\end{center}
\end{figure}

First, we benchmark the electron Bernstein modes ($s=e$). The result
is shown in Fig.\ref{fig:es3d_bern}(a), with parameter
$\omega_{pe}=2.5\omega_{ce}$. For the modes with frequency
$\omega<6\omega_c$, considering only the $N=10$-order Bessel
functions is accurate enough. The upper hybrid frequency calculated
at the cold limit is
$\omega_{UH}=\sqrt{\omega_c^2+\omega_p^2}=2.69$, which is consistent
with the matrix solution in the limit $k_{\perp}\rho_c\to0$.
Fig.\ref{fig:es3d_bern}(a) also agrees with Fig.9.8 in
Ref.\cite{Gurnett2005}. The corresponding ES1D3V particle-in-cell
(PIC) simulation (ion immobile, $k=k_\perp$) verification is also
shown in Fig.\ref{fig:es3d_bern}(b), where good agreement is
observed.

\subsubsection{Anisotropic instabilities}
\begin{figure}
\begin{center}
  \includegraphics[width=15cm]{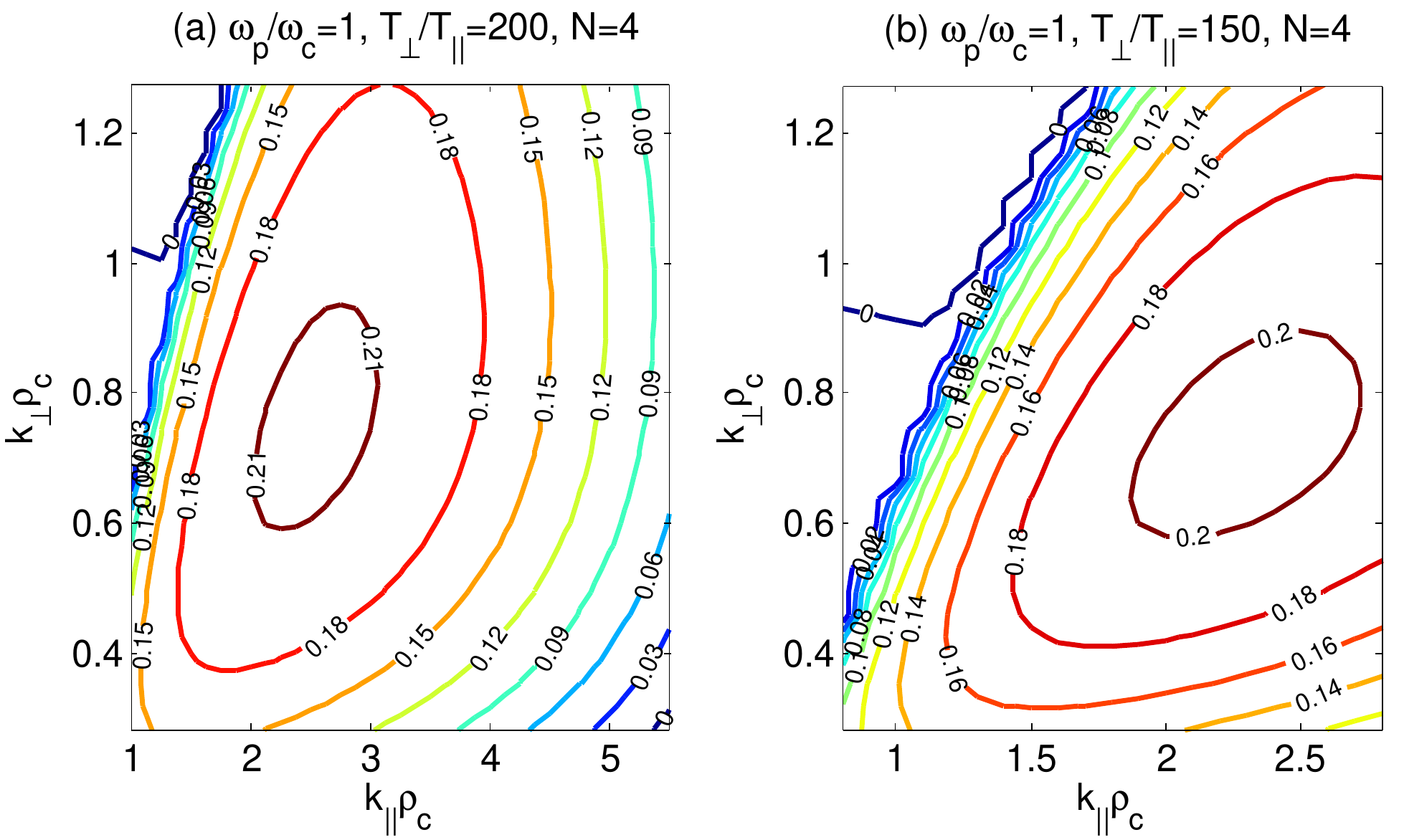}\\
  \caption{The anisotropic instabilities (growth rate $\gamma/\omega_c$) calculated from the Harris dispersion relation using the matrix
  method. The results agree with Ref.\cite{Gitomer1972}.}\label{fig:es3d_gitomer1972}
\end{center}
\end{figure}

Second, we benchmark the anisotropic instabilities with
Ref.\cite{Gitomer1972}. The contour plot of the growth rate
$\gamma/\omega_c$ is shown in Fig.\ref{fig:es3d_gitomer1972}, with
$\omega_{p}=\omega_{c}$ and $N=4$. The results agree with Fig.2 in
Ref.\cite{Gitomer1972}.

\section{Electromagnetic dispersion relation}\label{sec:em3d}
In the above section, we have shown that the matrix method can solve
the kinetic dispersion relations. In addition, the results are
accurate enough even if we used Pad\'e approximation to the $Z$
function, which gives us enough confidence with the approach to
extend its application further to the magnetized electromagnetic
(EM3D) dispersion relations, which has not been solved well using
conventional approaches.

\subsection{The dispersion relation}
The equilibrium distribution is still assumed to be drift
bi-Maxwellian as in Sec.\ref{sec:es3d}, and also ${\bm
B_0}=(0,0,B_0)$ and ${\bm k}=(k_x,0,k_z)$. The dispersion relation
can be derived as\cite{Stix1992}
\begin{equation}\label{eq:drkine}
    \left| \begin{array}{ccc}
    K_{xx}-\frac{c^2k^2}{\omega^2}\cos^2\theta & K_{xy} & K_{xz}+\frac{c^2k^2}{\omega^2}\sin\theta\cos\theta\\
    K_{yx} & K_{yy}-\frac{c^2k^2}{\omega^2} & K_{yz} \\
    K_{zx}+\frac{c^2k^2}{\omega^2}\sin\theta\cos\theta & K_{zy} & K_{zz}-\frac{c^2k^2}{\omega^2}\sin^2\theta
    \end{array}\right|=0,
\end{equation}
with ${\bm K}={\bm
I}+\sum_s\frac{\omega_{ps}^2}{\omega^2}\Big[\sum_n\big\{\zeta_0Z(\zeta_n)-(1-\frac{1}{\lambda_T})[1+\zeta_nZ(\zeta_n)\big\}{\bm
X_n}+2\eta_0^2\lambda_T{\bm L}\Big]$, where
\begin{equation}\label{eq:drXn}
    {\bm X_n}=\left( \begin{array}{ccc}
    n^2\Gamma_n/b & in\Gamma'_n & -(2\lambda_T)^{1/2}\eta_n\frac{n}{\alpha}\Gamma_n\\
    in\Gamma'_n & n^2/b\Gamma_n-2b\Gamma'_n & i(2\lambda_T)^{1/2}\eta_n\alpha\Gamma'_n \\
    -(2\lambda_T)^{1/2}\eta_n\frac{n}{\alpha}\Gamma_n & -i(2\lambda_T)^{1/2}\eta_n\alpha\Gamma'_n &
    2\lambda_T\eta_n^2\Gamma_n
    \end{array}\right),
\end{equation}
$\eta_n=\frac{\omega+n{\Omega}}{k_zv_{Tz}}$,
$\lambda_T=\frac{T_z}{T_\perp}$,
$b=(\frac{k_xv_{T\perp}}{\Omega})^{2}$,
$\alpha=\frac{k_xv_{T\perp}}{\Omega}$, $v_{T_z}^2=\frac{k_BT_z}{m}$,
$v_{T_\perp}^2=\frac{k_BT_\perp}{m}$ and the matrix components of
${\bm L}$ are all zero, except for $L_{zz}=1$.

\subsection{The linear transformation}
To seek an equivalent linear system, the Maxwell's equations
\begin{subequations} \label{eq:em3dmaxw}
\begin{eqnarray}
  & \partial_t {\bm E} = c^2\nabla\times{\bm B}-{\bm J}/\epsilon_0,\\
  & \partial_t {\bm B} = -\nabla\times{\bm E},
\end{eqnarray}
\end{subequations}
do not need be changed. We only need to seek a new linear system for
${\bm J}=\overleftrightarrow{\sigma}\cdot{\bm E}$. It is easy to
find that after $J$-pole expansion, the relations between $\bm J$
and $\bm E$ has the following form:
\begin{equation}\label{eq:JE}
    \left( \begin{array}{c}J_x \\ J_y \\ J_z\end{array}\right)
    =\left( \begin{array}{ccc}
    a_{11}+\sum_{snjm}\frac{b_{snjm11}}{\omega-c_{snjm11}} & a_{12}+\sum_{snjm}\frac{b_{snjm12}}{\omega-c_{snjm12}}
    & a_{13}+\sum_{snjm}\frac{b_{snjm13}}{\omega-c_{snjm13}}\\
    a_{21}+\sum_{snjm}\frac{b_{snjm21}}{\omega-c_{snjm21}} & a_{22}+\sum_{snjm}\frac{b_{snjm22}}{\omega-c_{snjm22}}
    & a_{23}+\sum_{snjm}\frac{b_{snjm23}}{\omega-c_{snjm23}} \\
    a_{31}+\sum_{snjm}\frac{b_{snjm31}}{\omega-c_{snjm31}} & a_{32}+\sum_{snjm}\frac{b_{snjm32}}{\omega-c_{snjm32}}
    & a_{33}+\sum_{snjm}\frac{b_{snjm33}}{\omega-c_{snjm33}}+d_{33}\omega
    \end{array}\right) \left( \begin{array}{c}E_x \\ E_y \\
    E_z\end{array}\right).
\end{equation}
Fortunately, noting the relations in $Z$ function ( $\sum_jb_j=-1$,
$\sum_jb_jc_j=0$ and $\sum_jb_jc_j^2=-1/2$) and in Bessel functions
[$\sum_{n=-\infty}^{\infty}I_n(b)=e^b$,
$\sum_{n=-\infty}^{\infty}nI_n(b)=0$,
$\sum_{n=-\infty}^{\infty}n^2I_n(b)=be^b$], we find that $a_{ij}=0$
($i,j=1,2,3$) and $d_{33}=0$. Eq.(\ref{eq:JE}) can be changed
further to
\begin{equation}\label{eq:JE1}
    \left( \begin{array}{c}J_x \\ J_y \\ J_z\end{array}\right)
    =-\left( \begin{array}{ccc}
    \frac{b_{11}}{\omega}+\sum_{snj}\frac{b_{snj11}}{\omega-c_{snj}} & \frac{b_{12}}{\omega}+\sum_{snj}\frac{b_{snj12}}{\omega-c_{snj}}
    & \frac{b_{13}}{\omega}+\sum_{snj}\frac{b_{snj13}}{\omega-c_{snj}}\\
    \frac{b_{21}}{\omega}+\sum_{snj}\frac{b_{snj21}}{\omega-c_{snj}} & \frac{b_{22}}{\omega}+\sum_{snj}\frac{b_{snj22}}{\omega-c_{snj}}
    & \frac{b_{23}}{\omega}+\sum_{snj}\frac{b_{snj23}}{\omega-c_{snj}} \\
    \frac{b_{31}}{\omega}+\sum_{snj}\frac{b_{snj31}}{\omega-c_{snj}} & \frac{b_{32}}{\omega}+\sum_{snj}\frac{b_{snj32}}{\omega-c_{snj}}
    & \frac{b_{33}}{\omega}+\sum_{snj}\frac{b_{snj33}}{\omega-c_{snj}}
    \end{array}\right) \left( \begin{array}{c}E_x \\ E_y \\
    E_z\end{array}\right).
\end{equation}

Combining Eqs.(\ref{eq:em3dmaxw}) and (\ref{eq:JE1}), the equivalent
linear system for (\ref{eq:drkine}) can be obtained as
\begin{equation}\label{eq:JEsystem}
    \left\{ \begin{array}{ccc}
    \omega v_{snjx} &=& c_{snj} v_{snjx} + b_{snj11} E_x + b_{snj12} E_y + b_{snj13} E_z, \\
    \omega j_x &=&  b_{11} E_x + b_{12} E_y + b_{13} E_z, \\
    J_x &=& j_x+\sum_{snj}v_{snjx}, \\
    \omega v_{snjy} &=& c_{snj} v_{snjy} + b_{snj21} E_x + b_{snj22} E_y + b_{snj23} E_z, \\
    \omega j_y &=&  b_{21} E_x + b_{22} E_y + b_{23} E_z, \\
    J_y &=& j_y+\sum_{snj}v_{snjy}, \\
    \omega v_{snjz} &=& c_{snj} v_{snjz} + b_{snj31} E_x + b_{snj32} E_y + b_{snj33} E_z, \\
    \omega j_z &=&  b_{31} E_x + b_{32} E_y + b_{33} E_z, \\
    J_z &=& j_z+\sum_{snj}v_{snjz}, \\
    \omega E_x &=& -c^2k_z B_y-J_x/\epsilon_0, \\
    \omega E_y &=& c^2k_z B_x -  c^2k_x B_z-J_y/\epsilon_0,\\
    \omega E_z &=& c^2k_x B_y-J_z/\epsilon_0, \\
    \omega B_x &=& k_z E_y, \\
    \omega B_y &=& -k_z E_x +  k_x E_z, \\
    \omega B_z &=& -k_z E_y,
    \end{array}\right.
\end{equation}
which yields a sparse matrix eigenvalue problem. The elements of the
eigenvector $(E_x, E_y, E_z, B_x, B_y, B_z)$ still represent the
original electric and magnetic fields. Thus, the polarization of the
solutions can also be obtained in a straightforward manner. The
dimension of the matrix is $NN=3\times (SNJ+1)+6=3\times
[S\times(2\times N+1)\times J+1]+6$. The coefficients are
\begin{equation}\label{eq:JEsystem}
    \left\{ \begin{array}{ccc}
    b_{snj11} &=&  \omega_{ps}^2b_j(1-k_zb_{j0}/c_{snj})n^2\Gamma_n/b_s,\\
    b_{11} &=&  \sum_{snj}\omega_{ps}^2b_j(k_zb_{j0}/c_{snj})n^2\Gamma_n/b_s,\\
    b_{snj12} &=&  \omega_{ps}^2b_j(1-k_zb_{j0}/c_{snj})in\Gamma'_n,\\
    b_{12} &=&  \sum_{snj}\omega_{ps}^2b_j(k_zb_{j0}/c_{snj})in\Gamma'_n,\\
    b_{snj21} =  -b_{snj12} &,&    b_{21} =  -b_{12},\\
    b_{snj22} &=&  \omega_{ps}^2b_j(1-k_zb_{j0}/c_{snj})(n^2\Gamma_n/b_s-2b_s\Gamma'_n),\\
    b_{22} &=&  \sum_{snj}\omega_{ps}^2b_j(k_zb_{j0}/c_{snj})(n^2\Gamma_n/b_s-2b_s\Gamma'_n),\\
    b_{snj13} &=&  \omega_{ps}^2b_j[c_j/\lambda_{Ts}-n\omega_{cs}b_{j0}/(c_{snj}v_{tzs})]\Gamma_n/b_s,\\
    b_{13} &=&  \sum_{snj}\omega_{ps}^2b_j[n\omega_{cs}b_{j0}/(c_{snj}v_{tzs})]\Gamma_n/b_s,\\
    b_{snj31} =  b_{snj13} &,& b_{31}=  b_{13},\\
    b_{snj23} &=&  -i\omega_{ps}^2b_j[c_j/\lambda_{Ts}-n\omega_{cs}b_{j0}/(c_{snj}v_{tzs})]\sqrt(2\lambda_{Ts})\Gamma'_nb_s,\\
    b_{23} &=&  -i\sum_{snj}\omega_{ps}^2b_j[n\omega_{cs}b_{j0}/(c_{snj}v_{tzs})]\sqrt(2\lambda_{Ts})\Gamma'_nb_s,\\
    b_{snj32} =  -b_{snj23} &,& b_{32} =  -b_{23},\\
    b_{snj33} &=&  \omega_{ps}^2b_j[(v_{s0}/v_{tzs}+c_j)c_j/\lambda_{Ts}-n\omega_{cs}b_{j0}(1+n\omega_{cs}/(c_{snj})v_{tzs}^2)/k_z]2\lambda_{Ts}\Gamma_n,\\
    b_{33} &=&  \sum_{snj}\omega_{ps}^2b_j[n^2b_{j0}/(c_{snj}v_{tzs}^2k_z)]2\lambda_{Ts}\Gamma_n,\\
    c_{snj} &=& k_zc_jv_{tzs}+k_zv_{s0}-n\omega_{cs},
    \end{array}\right.
\end{equation}
where $b_{j0}=v_{s0}+(1-1/\lambda_{Ts})c_jv_{tzs}$.

If $a_{ij}\neq0$, then the equivalent linear transformation is still
straightforward. However, the eigenmatrix will not be sparse (the
ES1D and ES3D eigenmatrices in Sec.\ref{sec:es} are not sparse, see
\ref{sec:spes} for the sparse ones). If $d_{33}\neq0$, then the
equivalent linear transformation will be more complicated. For our
purposes, we do not need to discuss these cases.

\section{Benchmarks and applications}\label{sec:bech}
The PDRK code is developed based on the above method. We now
benchmark this code and show some typical applications. Default
parameters for the succeeding cases are  $c^2=10^4$, $B_0=1$,
$m_e=1$, $q_e=-1$, $\epsilon_0=1$.

\subsection{Benchmark with fluid solver PDRF}
\begin{figure}
\begin{center}
  \includegraphics[width=15cm]{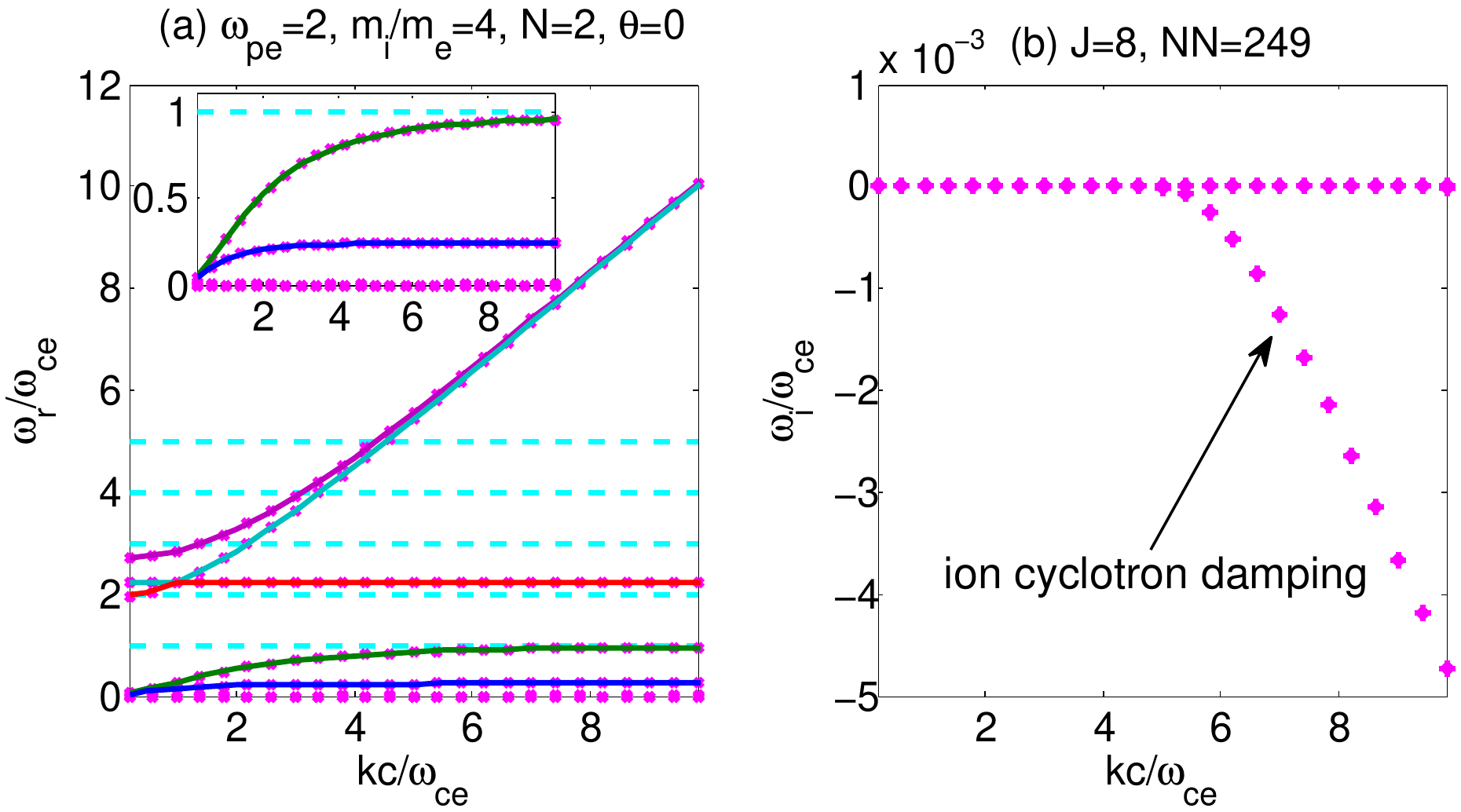}\\
  \caption{PDRK (dot) vs. PDRF (solid line), cold ($T_e=T_i=0.01$), parallel propagation.}\label{fig:pdrk_vs_pdrf_cold_para}
\end{center}
\end{figure}

First, we compare PDRK with the fluid solver PDRF\cite{Xie2014}.
Fig.\ref{fig:pdrk_vs_pdrf_cold_para} shows the results at the cold
limit with parallel propagation ($k=k_z$). In PDRF, we set
$T_e=T_i=0$; in PDRK, we set $T_e=T_i=0.01\ll1$. The real
frequencies in PDRK ($\omega^K$) and in PDRF ($\omega^F$) are almost
identical. However, the kinetic damping is not zero as in the fluid
framework, especially the cyclotron damping for ions, which is
apparent in Panel (b). This cyclotron damping is not predicted in
the fluid theory.

\begin{figure}
\begin{center}
  \includegraphics[width=15cm]{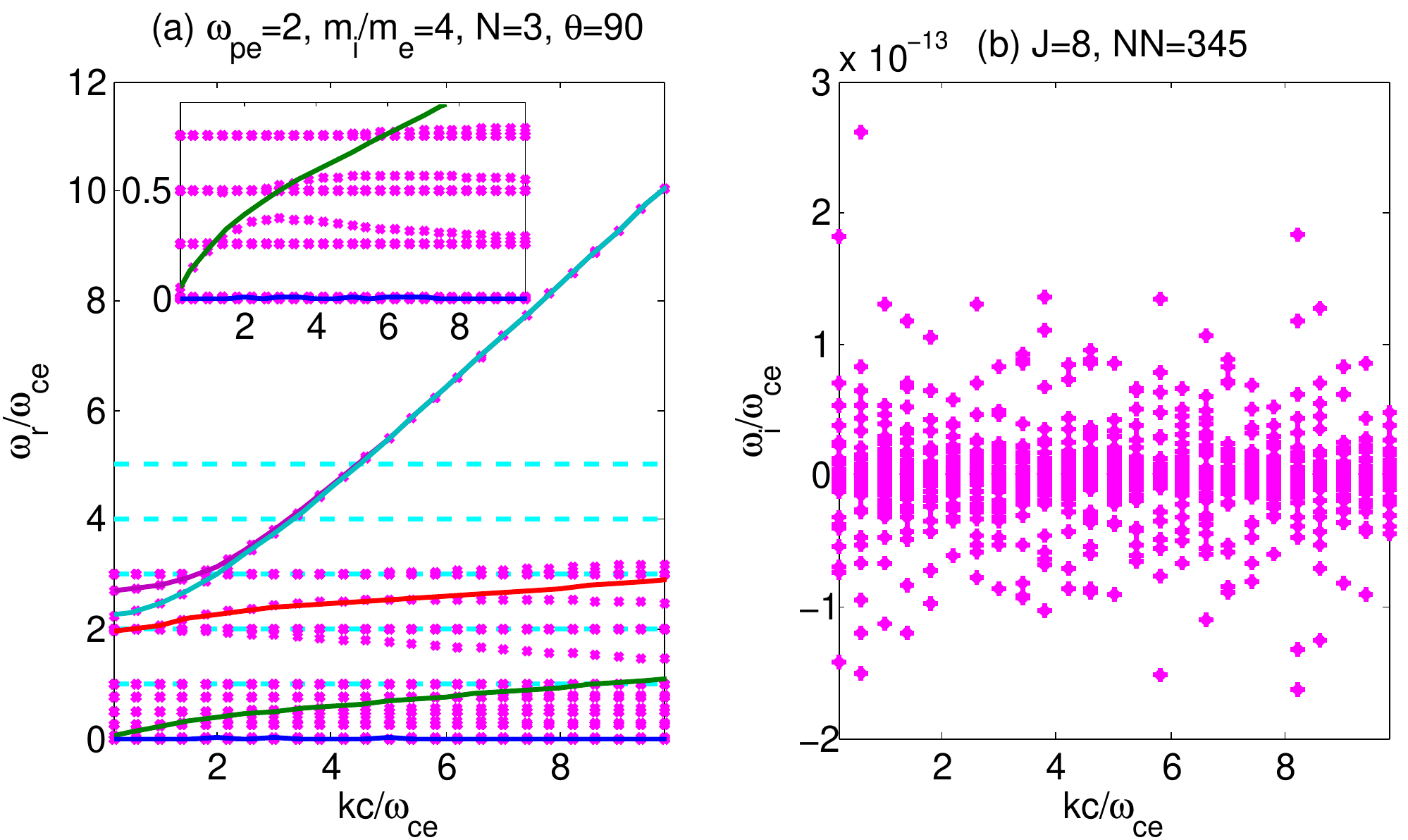}\\
  \caption{PDRK (dot) vs. PDRF (solid line), warm ($T_e=T_i=100$), perpendicular propagation.
  The positive $\gamma\simeq10^{-13}$ comes from numerical error of $J=8$.}\label{fig:pdrk_vs_pdrf_hot_perp}
\end{center}
\end{figure}

Fig.\ref{fig:pdrk_vs_pdrf_hot_perp} shows the results for warm
plasma with perpendicular propagation. We see that the fluid version
results are close to the kinetic version results at small $k$
($kc/\omega_{ce}<2$), but deviates at large $k$. This kinetic
correction (Bernstein modes) from the harmonics of the cyclotron
frequency is also not predicted in fluid theory.

A further test (Fig.\ref{fig:em3d_bern}) of the electron Bernstein
modes, which is quasi-electrostatic and makes use of the parameters
in Fig.\ref{fig:es3d_bern}, gives similar results between PDRK-EM3D
and PDRK-ES3D. Thus, for this step, PDRK-EM3D works well.

\subsection{Parallel propagation kinetic modes}
\begin{figure}
\begin{center}
  \includegraphics[width=13cm]{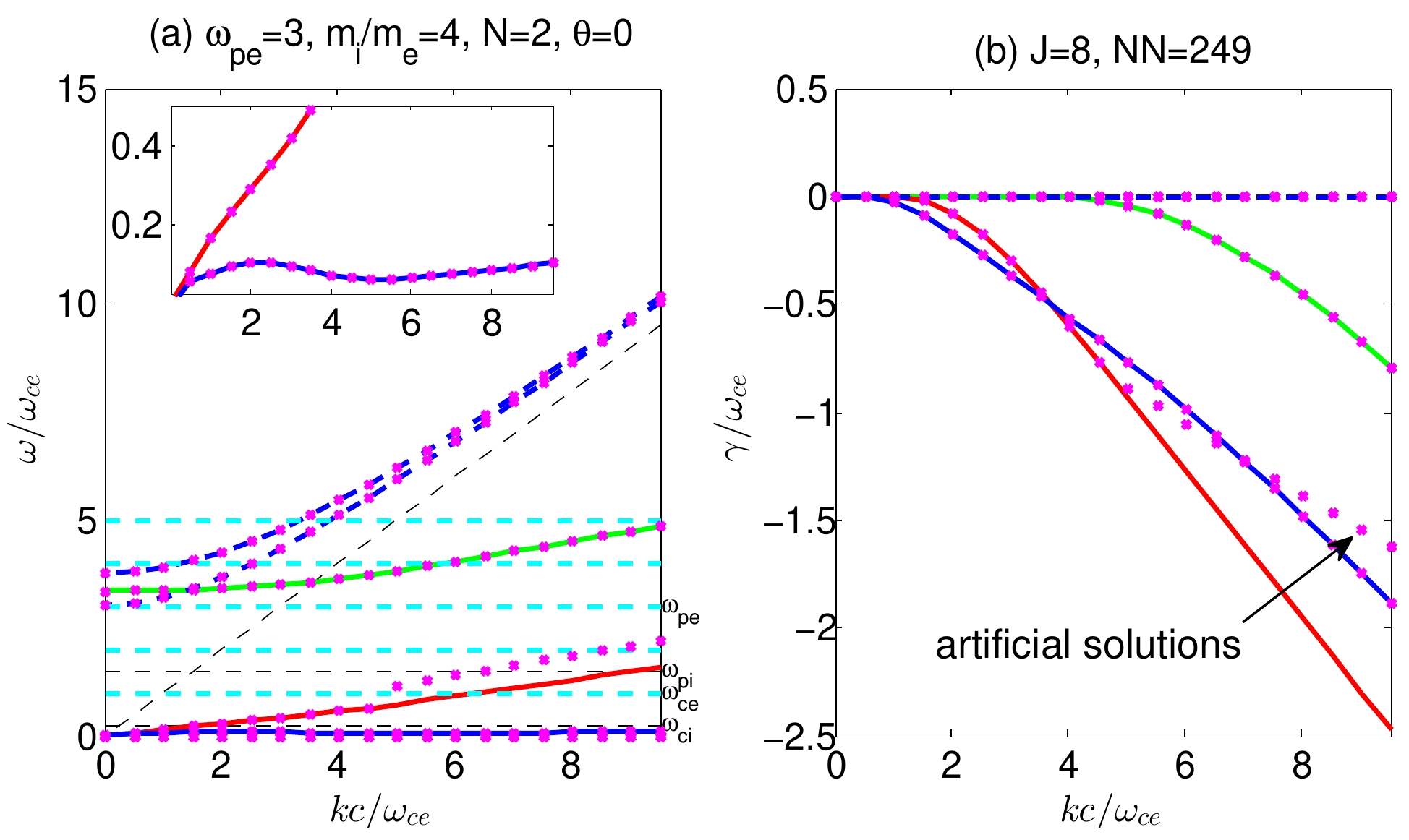}\\
  \caption{PDRK solutions (dot) vs. Z function solutions (solid and green dash lines), warm ($T_e=T_i=400$), parallel propagation. Heavily damped (both real and artificial) solutions
are not shown.}\label{fig:pdrk_vs_kntc_para}
\end{center}
\end{figure}

The kinetic dispersion relation for parallel propagation
modes\cite{Gurnett2005,Stix1992} is relatively simple to solve
because the effects of the higher-order cyclotron harmonics are
zero. One branch is the same as the ES1D dispersion relation
Eq.(\ref{eq:drkes1d}). The other two branches are given by
\begin{equation}\label{eq:em3ddr_para}
D(k,\omega) = 1 - \frac{k^2c^2}{\omega^2} + \sum_s \frac{\omega
_{ps}^2}{\omega k{v_{ts}}} Z\left( \frac{\omega \pm \omega_{cs}}
{kv_{ts}} \right) = 0.
\end{equation}
Eqs.(\ref{eq:drkes1d}) and (\ref{eq:em3ddr_para}) are solved by root
finding with the original $Z$ function\cite{Xie2014a} and comparing
with PDRK. A typical result is shown in
Fig.\ref{fig:pdrk_vs_kntc_para}. We find a good agreement between
the two methods. In addition, the ion and electron cyclotron damping
and the Landau damping are clearly shown. However, too many
extraneous solutions exist in the PDRK results. Most of the heavily
damped solutions are not shown in the figure. The solutions
represented by the red solid line ($\omega^R$) in the figure should
be real solutions. At large $k$ (e.g., $kc/\omega_{ce}>7$, where
PDRK solutions still agree with $\omega^R$ but not shown), the
damping rate of several artificial solutions are smaller than
$\omega^R$, which makes it difficult to separate the real and
artificial solutions directly.

To this step, PDRK-EM3D works well for $T_{s\parallel}=T_{s\perp}$
and $v_{s0}=0$. For the heavily damped solutions, keeping all the
interesting solutions while removing the artificial solutions is
usually not easy. Besides the heavily damped solutions, the
artificial solutions roughly satisfy $\omega_r-n\Omega_c\propto
k_\parallel$ and $\gamma\propto k_\parallel$ (come from the poles
$\zeta-c_j\to0$ of $J$-pole expansion). Therefore, this process can
also be used to remove some of the artificial solutions. Several of
the ES3D artificial solutions in Fig.\ref{fig:lhw_landau_esem} are
removed based on this property.

When a sparse matrix is not used, the computation time is around
$O(NN^\alpha)$ with $2<\alpha<3$ and the memory required is around
$O(NN^2)$. A typical personal computer with 4 GB memory can
calculate $NN$ up to $7000$ ($NN=7000$, $S=2$, $J=8$, give
$N\simeq60$ ) in minutes. Thus, for modes with frequency
$\omega<60\Omega_{ci}$, all the solutions in the system can be
obtained easily. When a sparse matrix is used, $NN$ can reach up to
$10^6$. Thus $N$ can be up to $10^4$. The standard sparse matrix
algorithm can solve one or several solutions around the initial
guess.

\subsection{Landau damping of lower hybrid wave}
\begin{figure}
\begin{center}
  \includegraphics[width=15cm]{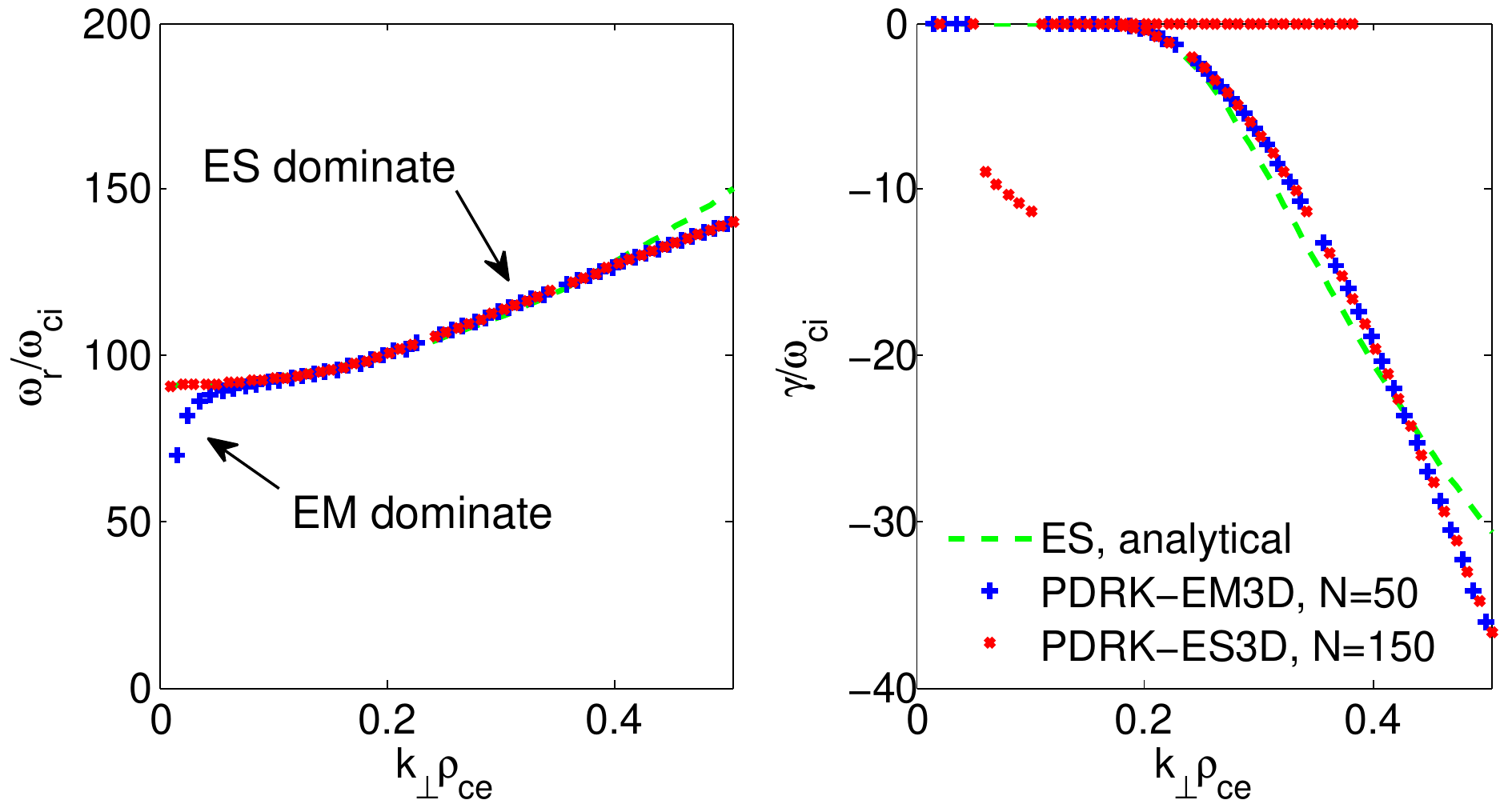}\\
  \caption{Landau damping of lower hybrid wave. Solutions from
  PDRK-ES3D (red, $N=150$), PDRK-EM3D (blue, $N=50$), and the analytical solution (dash green line) in Ref.\cite{Qi2013}.
  It took about 1 CPU hour to compute the data in this figure.}\label{fig:lhw_landau_esem}
\end{center}
\end{figure}

Now, we benchmark the Landau damping of lower hybrid wave (LHW)
using a real mass ratio $m_i/m_e=1836$, where large $N$ should be
used to make the solutions convergent. For the electrostatic case,
with $k^2\rho_e^2\ll1$, $\omega_{ci}\ll\omega\ll\omega_{pe}$ and
$k_\parallel/k\ll1$, the analytical solution
$\omega=(\omega_r,\gamma)$ for LHW can be found in
Ref.\cite{Qi2013}. We use the same parameters
($\omega_{pe}=\omega_{ce}$, $k_\parallel/k_\perp=0.066$, $T_e=T_i$)
as in the Fig.1 of Ref.\cite{Qi2013} for the benchmark because this
has also been verified by first-principle PIC simulations in that
paper. The results are shown in Fig.\ref{fig:lhw_landau_esem}, where
the electrostatic assumption works well for large $k$. For small $k$
($k_\perp\rho_{ce}<0.04$), the electromagnetic effects should be
included, which is consistent with the results on fluid frequency
and polarization in a previous study\cite{Xie2014}.

Note that several limits for the parameters have been used to obtain
the analytical solution. Similar limits have also been used for warm
EM LHW (see e.g., \cite{Verdon2009}).  Therefore, it is not
surprising that the analytical solution does not hold for large $k$
($k_\perp\rho_{ce}>0.4$) in the figure. For fusion (e.g.,
\cite{Bao2014}) or space studies, the approximate analytical
solution is not always valid. Thus, PDRK can serve as a numerical
tool for a wider range of parameters.

For this step, we have shown that PDRK-EM3D works well also for
$N\geq50$ by using a sparse matrix, although an initial guess is
required and the computational time is longer.

\subsection{Firehose and mirror modes}
Firehose and mirror modes are typical unstable modes driven by
pressure anisotropic $T_\parallel \neq T_\perp$. For cold electrons,
the approximate analytical kinetic dispersion relations for the
firehose mode is
$\omega^2=\omega_A^2[\frac{b_i}{1-\Gamma_0(b_i)}+\frac{\beta_{i\perp}-\beta_{i\parallel}}{2}]$.
For the mirror mode, it is
$\zeta_iZ(\zeta_i)=\frac{\eta_i}{\beta_{i\perp}\Gamma_1(b_i)}-(1-\eta_i)$,
with $\eta_i=\beta_{i\parallel}/\beta_{i\perp}$.

\begin{figure}
\begin{center}
  \includegraphics[width=15cm]{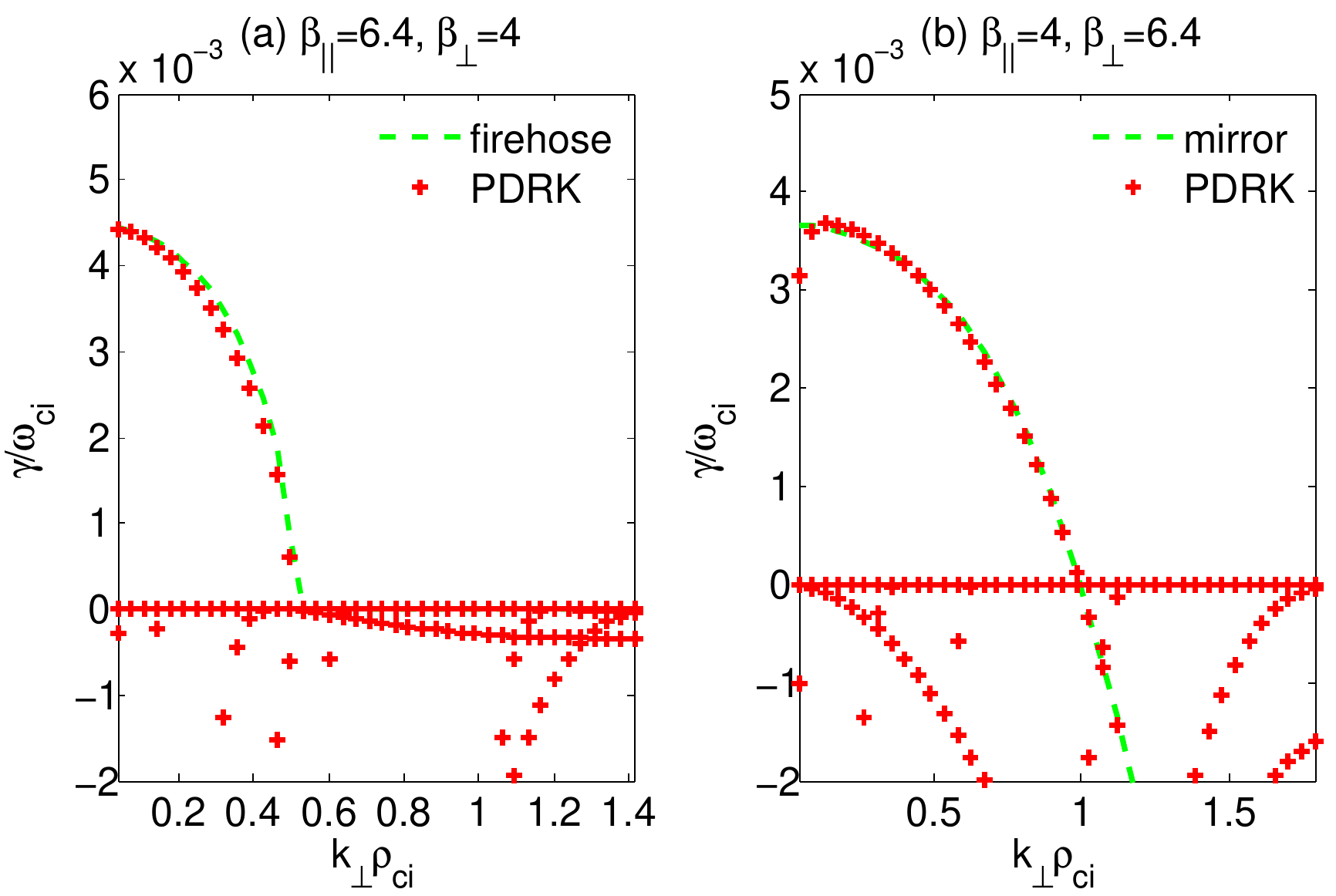}\\
  \caption{Growth rates for the firehose and mirror modes vs. $k_\perp\rho_{ci}$.
  The dashed green lines are analytical solutions.}\label{fig:fhmm}
\end{center}
\end{figure}

A typical result is shown in Fig.\ref{fig:fhmm}, where
$\omega_{pe}/\omega_{ce}=2$, $m_i/m_e=100$, $\omega_A=k_\parallel
v_A=0.01\omega_{ci}$ and $\beta_e=0.08$. The PDRK solutions agree
with the analytical solutions for both the firehose and mirror
modes. The small deviation is not surprising because the analytical
solutions are not accurate.

\subsection{Whistler beam mode}

\begin{figure}
\begin{center}
  \includegraphics[width=15cm]{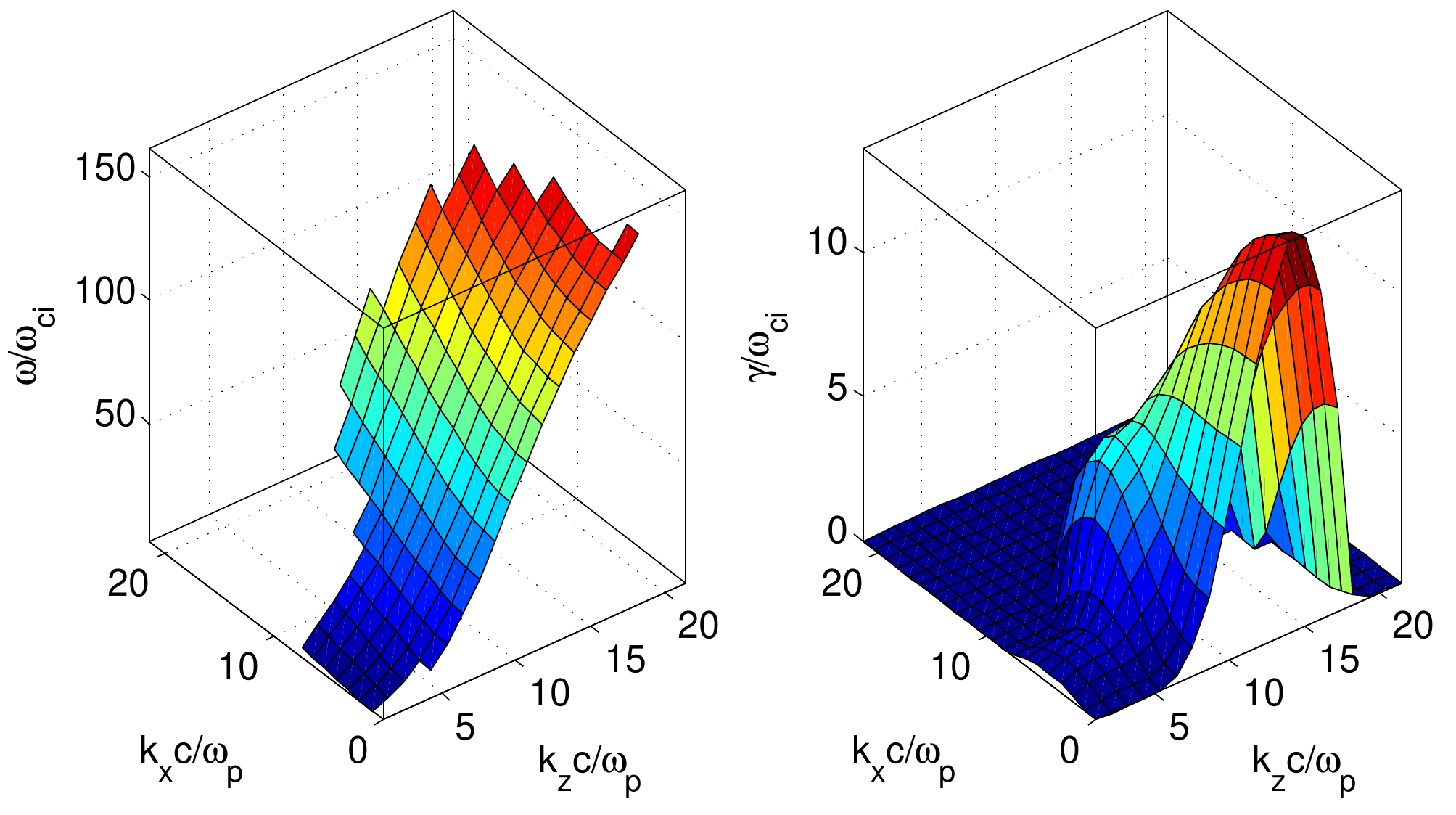}\\
  \caption{Electromagnetic whistler beam instability. The real
  frequency $\omega$ is only shown for unstable ($\gamma>0$) solutions.
  The parallel propagation ($k=k_\parallel$) results are similar to Fig.8.8
of Ref.\cite{Gary1993}. $N=3$ is used for this
calculation.}\label{fig:whistler_gary1993}
\end{center}
\end{figure}

The beam $v_{s0}\neq0$ can also drive instabilities. We benchmark
the whistler beam mode here. The parameters are similar to Fig.8.8
of Ref.\cite{Gary1993}, with $s=b,c,i$, $m_i/m_e=1836$, $n_i=1.0e4$,
$n_b=0.1n_i$, $n_c=0.9n_i$,  $T_c=T_i=T_b/10=0.5556$ and
$v_{b0}=-9v_{c0}=2.108$, which yield $\omega_{pe}=100\omega_{ce}$,
$\beta_c=1.0$ and $v_{b0}=2.0v_{tc}$. The $\omega$ and $\gamma$ vs.
$(k_z,k_x)$ results are shown in Fig.\ref{fig:whistler_gary1993}.
The most unstable mode is the parallel propagation mode (
$k=k_\parallel$), which is consistent with Gary's
conclusion\cite{Gary1993}.

\subsection{New anomalous Doppler shift}
With PDRK, it was the first time that we can see a complete picture
of the waves and instabilities in a kinetic system. New modes which
are unknown in previous studies, may now be found. Several examples
of ¡®new¡¯ modes have been found by PDRK. We show one of them here,
namely, a new anomalous Doppler effect.

\begin{figure}
\begin{center}
  \includegraphics[width=15cm]{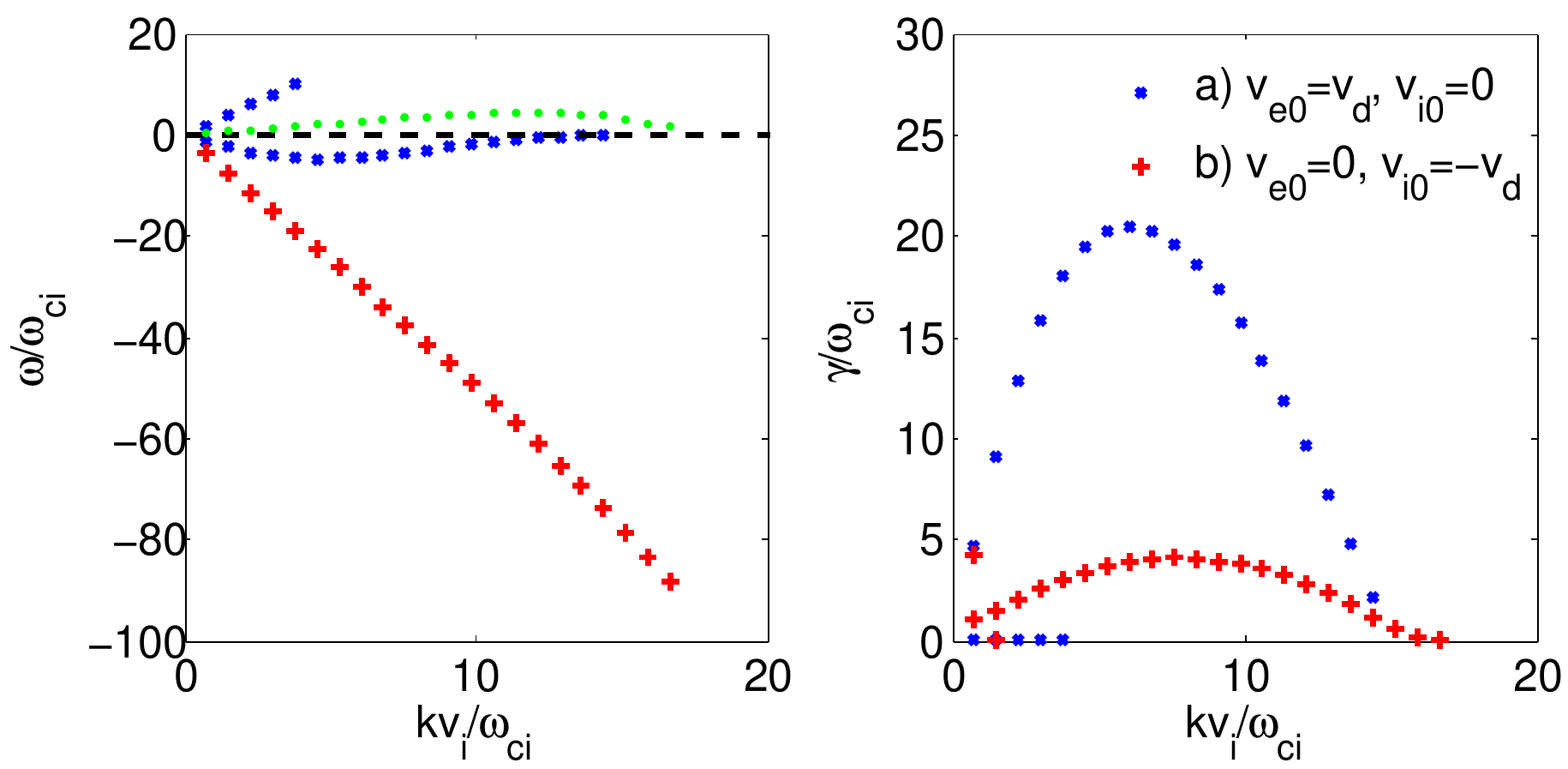}\\
  \caption{New modes found by PDRK-EM3D: the Doppler asymmetry of electron and ion beams in kinetic
non-relativistic plasmas. Blue `$\times$' for a); red `+' for b);
green dot is $\omega=\omega^b-k_\parallel v_d$. }\label{fig:doppler}
\end{center}
\end{figure}

The Lorentz Doppler shift for relativistic cold fluid plasma has
been verified by PDRF\cite{Xie2014}. Here, we are interested in the
Doppler asymmetry of the electron and ion beams in kinetic
non-relativistic plasmas. The parameters
($\omega_{pe}=100\omega_{ce}$, $m_i=1836m_e$, $T_i=T_e=0.01$,
$\theta=1.5528=88.97^\circ$ and $v_d=0.99$) are taken similar to
those in Fig.1 of Ref.\cite{Verdon2011} for instability of the lower
hybrid-like waves driven by parallel current. The current is taken
by electron beam in Ref.\cite{Verdon2011}. We also consider an ion
beam and solve the dispersion relations for the following two cases:
(a) $v_{e0}=v_d$, $v_{i0}=0$; (b) $v_{e0}=0$, $v_{i0}=-v_d$. Here,
the thermal velocity $v_{ts}$ and drift velocity $v_d$ are all
non-relativistic, i.e., $v_{ts}, v_d <0.01c \ll c$. If the system is
Galilean invariant, then the solution $\omega^a$ for (a) and
$\omega^b$ for (b) should satisfy $\omega^a=\omega^b-k_\parallel
v_d$. The foregoing also means that the growth rate will not change
($\gamma^a=\gamma^b$) for the same $k$.

Evidently, the ES dispersion relations (\ref{eq:drkes1d}) and
(\ref{eq:drkes3d}) are Galilean types, i.e.,
$\omega^a=\omega^b-k_\parallel v_d$. In common understanding, the
EM3D dispersion relation (\ref{eq:drkine}) should also be Galilean
when $v_{ts}, v_d \ll c$. However, as the results show in
Fig.\ref{fig:doppler}, we find that the Galilean Doppler effect is
not present for both real frequency and growth rate, i.e.,
$\omega^a\neq\omega^b-k_\parallel v_d$ and $\gamma^a\neq\gamma^b$.
Detailed discussion of the physics behind this interesting result is
not within the scope of the present work and may be explored
further. The purpose of the result shown here is to demonstrate that
PDRK can be useful and effective in revealing new modes.

\subsection{Dispersion surface}

\begin{figure}
\begin{center}
  \includegraphics[width=15cm]{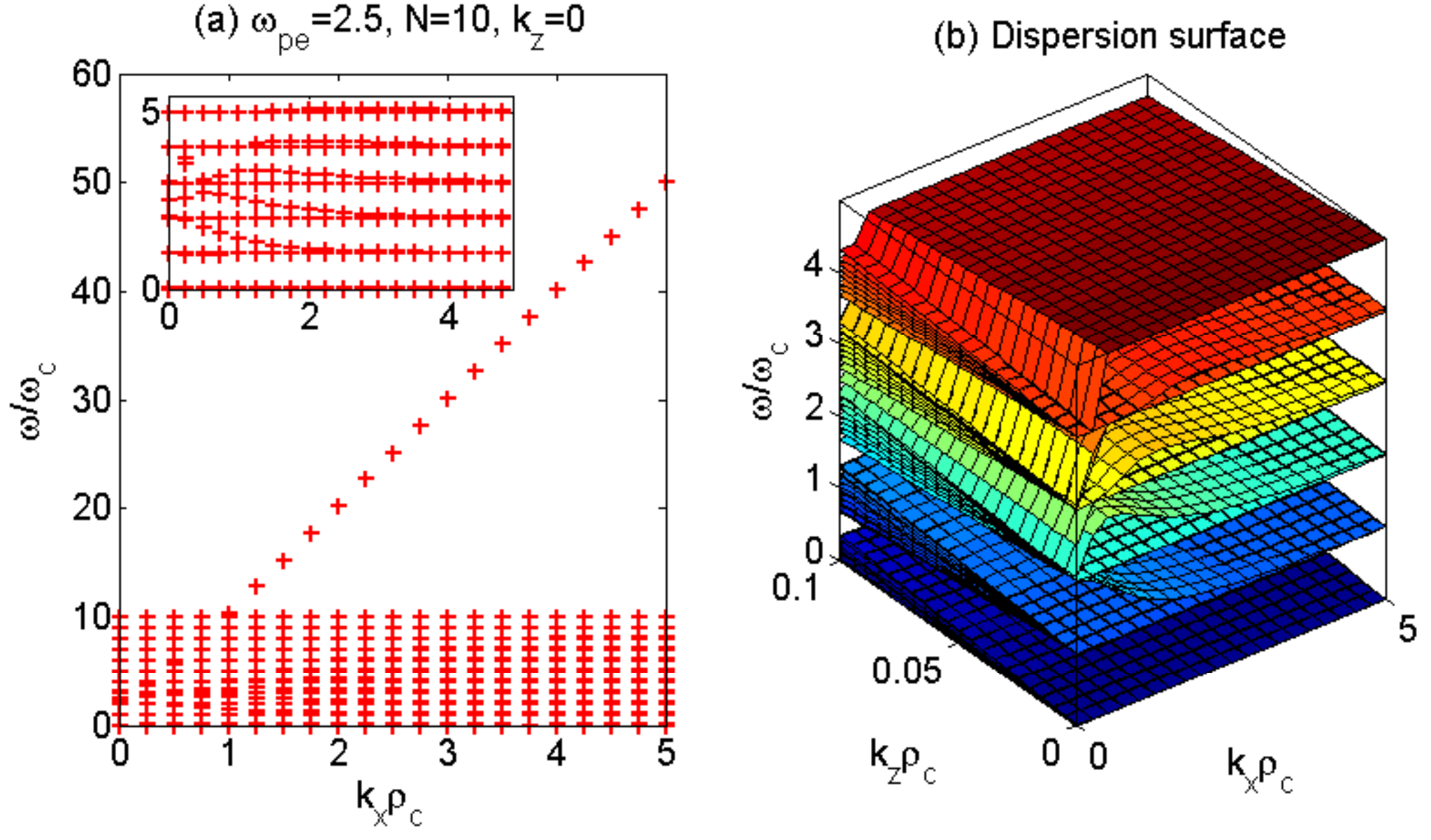}\\
  \caption{Dispersion surface (b) from PDRK-EM3D, using the EBW parameters
  in Fig.\ref{fig:es3d_bern} and $c^2=10^2$. The $\omega$ vs. $k_\perp$ (a) result is close to the
  ES3D result in Fig.\ref{fig:es3d_bern}, which confirms that EBW is (quasi-) electrostatic.}\label{fig:em3d_bern}
\end{center}
\end{figure}

The 2D structure of $\omega$ vs. $(k_x,k_z)$ (dispersion
surface\cite{Andre1985}) is shown in Fig.\ref{fig:em3d_bern} for
electron Bernstein wave (EBW). This type of figure is helpful in
displaying the fine structure of the dispersion relations in
$(k_\perp,k_\parallel)$ space and in revealing the relations among
different modes. It is clearly shown in Panel (b) that the solutions
are separated by cyclotron frequencies, i.e., the solution
$n\omega_c<\omega<(n+1)\omega_c$ ($n=0,1,2,...$) exists for any $k$.
In Fig.\ref{fig:em3d_bern}, we only keep $N=10$, and both real and
artificial solutions are shown. To see the fine structure of the
real solutions more clearly, further processing is required to
remove the artificial solutions, which is the main disadvantage of
the present version of PDRK.

\subsection{Others}
In the above benchmarks, no apparent numerical problems are found.
However, this does not mean that we can apply PDRK for all cases
because only approximations of $Z$ function are used. In
WHAMP\cite{Ronnmark1982}, the $Z$ function is also approximated but
$J$-pole expansion is used. A further approximation is needed for
the Bessel function summation. Thus, in principle, PDRK-EM3D will
give more accurate results than WHAMP. Similar issues regarding the
validity of Pad\'e approximation for $Z$ is discussed in detail in
the WHAMP report\cite{Ronnmark1982}. Based on our results, the error
for $J=8$ is less than $10^{-4}$, which may bring some artificial
growing modes. If the same solution also exists for other $J$ (e.g.,
$J=4, 12$), it is more likely to be a real solution. Otherwise, care
should be exercised in treating this solution. We can distinguish
real and artificial solutions by using different $J$. The artificial
solutions change when $J$ changes. By contrast, the real solutions
do not change that much.

\section{Summary and discussion}\label{sec:summ}
A general kinetic plasma dispersion relation solver, PDRK (three
versions are included at present: ES1D, ES3D, EM3D), is developed,
where the equilibrium distribution function is assumed to be drift
bi-Maxwellian. For other non-Maxwellian distribution functions, the
$J$-pole expansion (\ref{sec:jexpan}) of the corresponding new $Z$
functions\cite{Xie2013} should be obtained first. Note that the
relativistic effect (e.g., \cite{Bret2010,Hao2012,Timofeev2013}) is
not included in the present study as this would make the solution
more complicated. However, in principle, it can also be treated
using Pad\'e approximation\cite{Hao2012}. Although PDRK is more
accurate than PDRF, the latter is still advantageous in some cases
because it can handle more configurations, such as relativistic
systems, local non-uniform systems, and systems where collisions are
considered. In addition, it does not produce artificial solutions.
For practical applications, one can use PDRF to obtain rough
solutions, and then use these to provide initial guesses for PDRK or
use them for assistance in removing the artificial solutions in
PDRK. Besides the multi-fluid model, PDRK also provides a tool to
check the validity of other reduced models, such as
Darwin\cite{Xie2014a} and gyro-kinetic\cite{Lin2005,Howes2006}
models.

For systems with small $N$ (e.g., $N<60$ for two species) or
unstable modes, PDRK works excellently and is applicable to most
cases used. For large $N$ (e.g., $N>60$), especially in studying the
effect of $n\Omega_c$ to the modes (e.g., LHW), the performance of
PDRK is limited mainly by the computational time and memory.
However, this concern may be remedied by using sparse matrices.
Further optimization is possible. For example, we do not need to
treat $N$ equally for each species, e.g., for LHW, we can use large
$N_i$ but small $N_e$. The main disadvantage of PDRK is that the
artificial solutions originate from the poor approximation for
strongly damped modes.

Compared with conventional solvers, the PDRK solver is fast and can
give all solutions. Therefore, no important solutions are missed. It
is also free from convergence problems. Hence, this solver can find
wide applications in space, astrophysical, laser, and laboratory
plasma studies.

\section{Acknowledgements}
The work is supported by the National Magnetic Confinement Fusion
Science Program under Grant No. 2011GB105001 and 2013GB111000, China
NSFC under Grant No. 91130031, the Recruitment Program of Global
Youth Experts.

\appendix

\section{Arbitrary $J$-pole expansion}\label{sec:jexpan}
The $J$-pole expansion coefficients $b_j$ and $c_j$ are provided
only for small $J$ in literature. Here, based on the study of
Ronnmark\cite{Ronnmark1982}, we develop a scheme to calculate the
numerical coefficients for any $J$. This is possible because we do
not need the analytical expressions. The $J$-pole expansion is
\begin{equation}\label{eq:ZA1}
    Z(\zeta)\simeq
    Z_A^J(\zeta)=\frac{\sum_{k=0}^{J-1}p_k\zeta^k}{q_0+\sum_{k=1}^{J}q_k\zeta^k},
\end{equation}
with $q_0=1$, should be matched with the following two-side
approximation
\begin{equation}\label{eq:ZA}
    Z(\zeta)\simeq\left\{ \begin{array}{lll}
    \sum_{k=0}^\infty a_k\zeta^k &\simeq i\sqrt{\pi}e^{-\zeta^2}-\zeta\sum_{n=0}^\infty(-\zeta^2)^{n}\frac{\Gamma(1/2)}{\Gamma(n+3/2)}, & \zeta\to0\\
    \sum_{k=0}^\infty a_{-k}\zeta^{-k} &\simeq i\sigma\sqrt{\pi}e^{-\zeta^2}-\sum_{n=0}^\infty\frac{\Gamma(n+1/2)}{\Gamma(1/2)\zeta^{2n+1}}, & \zeta\to\infty
    \end{array}\right.
\end{equation}
where
\begin{equation}\label{eq:Zsigma}
    \sigma=\left\{ \begin{array}{lll}
    0 &, & {\rm IM}(\zeta)>0,\\
    1 &, & {\rm IM}(\zeta)=0,\\
    2 &, & {\rm IM}(\zeta)<0,
    \end{array}\right.
\end{equation}
and $\Gamma$ is Euler's Gamma function. A further expansion is
$e^{-\zeta^2}=\sum_{n=0}^\infty\frac{\zeta^{2n}}{n!}$. However,
$i\sigma\sqrt{\pi}e^{-\zeta^2}$ is omitted, which does not match
well for the range $y<\sqrt{\pi}x^2e^{-x^2}$ when $x\gg1$. The
system of equations to be solved are
\begin{subequations}\label{eq:ZJpq}
\begin{eqnarray}
  & p_j = \sum_{k=0}^ja_kq_{j-k},  1\leq j \leq I\label{eq:ZJp}\\
  & p_{L-j} = \sum_{k=0}^ja_{-k}q_{L+k-j},   1\leq j \leq K \label{eq:ZJq}
\end{eqnarray}
\end{subequations}
where $I+K=2J$, and $p_j=0$  for $j>J-1$ and $j<0$, and $q_j=0$ for
$j>J$ and $j<0$. Thus $2J$ equations determine $2J$ coefficients
$p_j$ and $q_j$ in (\ref{eq:ZA1}). The derivation of (\ref{eq:ZJpq})
is similar to that of Eqs.(III-5) and (III-7) in
Ronnmark\cite{Ronnmark1982}. Eqs.(\ref{eq:ZJpq}) are solved using
matrix inversion. The `residue()' function in MATLAB is used to
calculate $b_j$ and $c_j$ in (\ref{eq:NZ}) from (\ref{eq:ZA1}). The
results for $J=12$ using $I=16$ equations of (\ref{eq:ZJp}) and
$K=8$ equations of (\ref{eq:ZJq}) are given in
Table.\ref{tab:NpoleZ}.

Usually, a large $J$ gives better approximations. However, this is
not always the case. Test should be made before using them.
Moreover, the truncated error when using double precision data can
accumulate to $10^{-11}$.

Calculating the $J$-pole expansions for other equilibrium
distribution functions\cite{Xie2013} is also straightforward. We
merely replace the coefficients $a_k$ and $a_{-k}$ in (\ref{eq:ZA}).

\section{Equivalent sparse matrix for ES1D system}\label{sec:spes}
As mentioned, the equivalent matrix from Eq.(\ref{eq:fpeq}) for ES1D
system is not sparse. An equivalent sparse matrix for ES1D system
can be constructed as following:
\begin{subequations} \label{eq:spfpeq}
\begin{eqnarray}
  & \omega n_{sj} = c_{sj}n_{sj}+b_{sj}E,\\
  & \omega E = -\sum_{sj}c_{sj}n_{sj}-\sum_{sj}b_{sj}E.\label{eq:spfpeq_b}
\end{eqnarray}
\end{subequations}
This is similar by changing the ES1D Vlasov-Poisson system to the
ES1D Vlasov-Ampere system\cite{Xie2013a}. Eq.(\ref{eq:spfpeq_b}) can
be further simplified to be $\omega E = -\sum_{sj}c_{sj}n_{sj}$,
because $\sum_{sj}b_{sj}=0$. The ES3D matrix in Sec.\ref{sec:es3d}
can be changed to sparse matrix in a similar manner.

\section{PDRK User Manual}\label{sec:manu}
The structure of PDRK is similar to that of PDRF, i.e., it contains
two files: the main program ``pdrk.m" and the input data file
``pdrk.in". The input file has the following structure
\begin {verbatim}
qs      ms        ns       Tzs       Tps      vs0
-1.0    1.0       4.0      1.0       1.0      0.0
1.0     4.0       4.0      1.0       1.0      0.0
\end {verbatim}
More species can be added directly to new lines. Implementing
``pdrk.m" in other languages (e.g., Fortran, C/C++, Python) is also
straightforward.

\end{document}